\documentclass[twocolumn,showpacs,preprintnumbers,amsmath,amssymb]{revtex4}
\usepackage{amssymb}
\usepackage{graphicx}
\usepackage{dcolumn}
\usepackage{bm}

\begin{document}
\title[]{The optical selection rules of a graphene quantum dot in external electric fields}

\author{Qing-Rui Dong}
\address{School of Physics and Electronics, Shandong Normal University, Jinan, Shandong, 250014, People's Republic of China}
\author{Chun-Xiang Liu}
\address{School of Physics and Electronics, Shandong Normal University, Jinan, Shandong, 250014, People's Republic of China}

\begin{abstract}
We study theoretically the single-electron triangular zigzag graphene quantum dot in three typical in-plane electric fields.
The far-infrared absorption spectra of the dot are calculated by the tight-binding method and then the optical selection rules are identified by contrast with the corresponding energy spectra.
Our result shows that there exist the remarkable optical selection rules due to the $C3$ symmetry of the dot.
When the electric field possesses also the $C3$ symmetry, there are only two absorption peaks in the absorption spectra.
As the $C3$ symmetry of the system is damaged by the electric fields, both the intensity of the strongest peak and the number of the forbidden transitions decrease gradually.
Moreover, the polarization causes the decrease of the peak intensities and even new forbidden transitions.
Our findings may be useful for the application of graphene quantum dots to electronic and optoelectronic devices.
\end{abstract}

\maketitle

\section{Introduction}
Graphene, a single layer of carbon atoms arranged in a two-dimensional honeycomb lattice, was first successfully fabricated in 2004\cite{Novoselov04}. Due to the exceptional properties, such as massless carrier behavior\cite{katsnelson2006chiral}, high carrier mobility at room temperature\cite{bolotin2008ultrahigh}, superior thermal conductivity\cite{Balandin2008}, extremely high tensile strength\cite{lee2008measurement} and high transparency to incident light over a large wavelength range\cite{nair2008fine}, graphene has attracted enormous research interest and exhibited great application potential in next-generation electronics\cite{Novoselov12} and optoelectronics\cite{Xia14}. Much of the current understanding of the electronic properties of graphene has
been reviewed by Castro-Neto\cite{neto2009electronic}, transport properties by Das Sarma\cite{sarma2011electronic}
and many-body effects by Kotov\cite{kotov2012electron}.
However, a gap has to be induced in the gapless graphene for its real applications in electronic devices.
For this purpose, graphene quantum dots (GQDs) have been proposed as one of the most promising kinds of graphene nanostructures\cite{gucclu2014graphene}.
GQDs exhibit the unique electronic, spin and optical properties, which allow them hold great application potential in electronics and optoelectronics such as super capacitor\cite{liu2013superior}, flash memory\cite{joo2014graphene}, photodetector\cite{kim2014high} and phototransistor\cite{konstantatos2012hybrid}.
On the other hand, with recent developments of fabrication techniques, it is possible to cut accurately the bulk graphene into different sizes and shapes, such as hexagonal zigzag quantum dots, hexagonal armchair quantum dots, triangular zigzag quantum dots and triangular armchair quantum dots\cite{bacon2014graphene}.

Further applications of GQDs require a thorough knowledge of their electronic properties.
The electronic and magnetic properties of GQDs depend strongly on their shapes and edges\cite{Gu09,abergel2010properties,Potasz12}.
Moreover, for zigzag GQDs, especially triangular GQDs (TGQDs), there appears a shell of degenerate states at the Dirac points and the degeneracy is proportional to the edge size\cite{potasz2010zero,JZT2014}.
As a result of the degenerate zero-energy band, magnetism arising in graphene nanostructures (nanoflakes, quantum dots and nanoribbons) has recently collected rich literature\cite{sun2017magnetism,basak2016optical,hawrylak2016carbononics}.
The key feature for device application of GQDs is the ability to manipulate their electronic structures.
Therefore, one of the flourishing fields of exploration is the influence of external fields on the degenerate zero-energy band\cite{farghadan2014Iran}.
The electronic structure and magnetization relating to the zero-energy band can be manipulated electrically\cite{Chen10,Ma12PRB,Dong13}, optically\cite{Gu13} and magnetically\cite{szalowski2015Poland,Peeters2017exciton}.
In particular, the electrical manipulation of the zero-energy band of such GQDs is quite important for the operation of related devices, since it is easier to generate the potential field through local gate electrodes than the optical or magnetic field.
However, it is rather rare to study the influence of electric fields on the optical properties relating to the zero-energy band\cite{abdelsalam2016electro}.

In this paper, we concentrate on the effects of three typical in-plane electric fields on the far-infrared
(FIR) absorption spectra of a TGQD.
Our result shows that there exist the remarkable selection rules in the FIR spectra due to the $C3$ symmetry of the dot.
When the electric field possesses also the $C3$ symmetry, there are only two absorption peaks in the FIR spectra.
As the $C3$ symmetry of the system is damaged by the electric fields, both the intensity of the strongest peak and the number of the forbidden transitions decrease gradually.
Our findings may be useful for the application of GQDs to electronic and optoelectronic devices.
\section{Model and method}
In order to study the FIR spectrum of a single-electron GQD, we propose a scheme for the single-electron system.
The theoretical basis of this scheme is the Coulomb blockade effect in GQDs\cite{guttinger2012transport}.
The number of electrons in the dot is determined by the condition that the chemical potential of the dot is less than that of the leads (source and drain)\cite{Wiel2003RevModPhys}.
The chemical potential of the dot $\mu(N)$ is defined as $\mu(N)=E_G (N)-E_G (N-1)$, where $E_G (N)$ is the ground-state energy of the $N$-electron system.
For a single-electron system, more simply, $\mu(1)=E_G (1)$.
In other words, the single-electron system can be obtained if the ground-state energy of the system is slightly lower than the chemical potential of the leads.

The low-energy electronic structure of a GQD subjected to an in-plane electric field can be calculated by means of the tight-binding method\cite{Chen10,JZT2014}.
In the low-energy range, the tight-binding Hamiltonian with the nearest-neighbor approximation proves to give the same accuracy as first-principle calculations\cite{Abergel10}.
The Hamiltonian equation of the system is $H|\Psi(\textbf{r})\rangle=E|\Psi(\textbf{r})\rangle$ and the tight-binding Hamiltonian with the nearest-neighbor approximation is\cite{Ma12}
\begin{equation}\label{eqn:1}
H=\sum_{n}{(\varepsilon_{n}+U_{n})C^{+}_{n}C_{n}}+\sum_{<n,m>}{t_{n,m}C^{+}_{n}C_{m}},
\end{equation}
where $n$, $m$ denote the sites of carbon atoms in graphene, $\varepsilon_{n}$ is the on-site energy of the site $n$, $U_{n}$ is the electrostatic potential of the site $n$ obtained by solving a Laplace equation, $t_{n,m}$ is the hopping energy and $C^{+}_{n}$ ($C_{n}$) is the creation (annihilation) operator of an electron at the site $n$. The summation $<n,m>$ is taken over all nearest neighboring sites.
Due to the homogeneous geometrical configuration, the on-site energy and the hopping energy may be taken as $\varepsilon_{n}$ = 0 and $t_{n,m}$ = 2.7 eV.

Using the Fermi golden rule with the electric-dipole approximation for the perturbing unpolarized light, the transition probability from the ground state to the $l$th excited state can be calculated as\cite{dong2007,abdelsalam2016electro}
\begin{equation}\label{eqn:2}
A_l\varpropto|\langle \Psi_{l}|\textbf{r}|\Psi_{0}\rangle
|^{2}\delta(E_{l}-E_{0}-\hbar \omega),
\end{equation}
In addition to that, one selected spectrum can be decomposed to $x$ and $y$ polarization,
\begin{equation}\label{eqn:3}
\left\{ \begin{gathered}
  A_l^x\varpropto|\langle \Psi_{l}|x|\Psi_{0}\rangle
|^{2}\delta(E_{l}-E_{0}-\hbar \omega) \\
  A_l^y\varpropto|\langle \Psi_{l}|y|\Psi_{0}\rangle
|^{2}\delta(E_{l}-E_{0}-\hbar \omega)  \\
\end{gathered}  \right.
\end{equation}
According to the irreducible theory of the symmetry group\cite{elliott1979symmetry}, symmetry leads to selection rules or forbidden transitions.
For the same system, the transition matrix element for the polarized light $A_l^x$ or $A_l^y$ is a component of $A_l$.
Thus, the polarization may cause the decrease of the transition probabilities and even forbidden transitions.

\begin{figure}[htp]
\centering
\includegraphics[height=13cm]{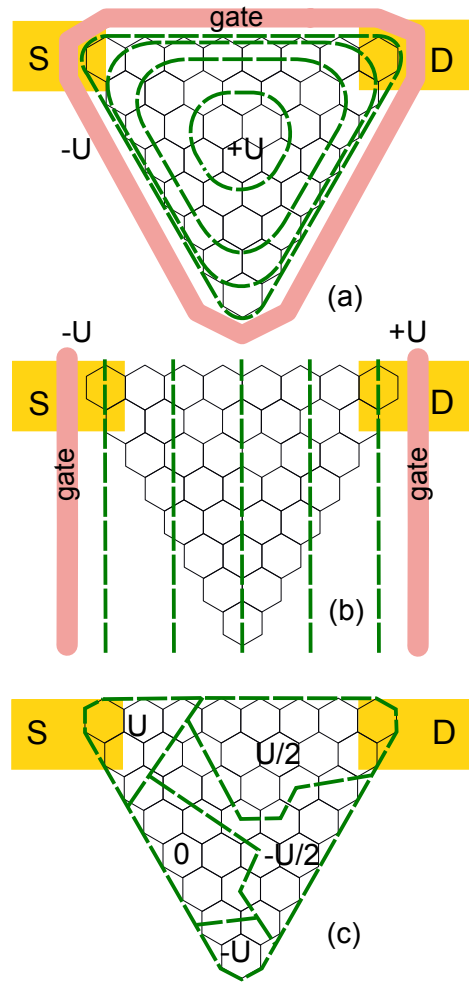}\caption{\label{fig:EF}
The electric fields applied to a TGQD ($N_{s}$ = 8). (a) The triangular electric field $EF1$ with a $C3$ rotation symmetry, where two gates with electrostatic potentials $\pm U$ are applied outside and bottom of the dot. (b) The uniform electric field $EF2$, where two gates with electrostatic potentials $\pm U$ are applied to the left and right of the dot. (c) The random electric field $EF3$ which presents randomly an imaginary potential distribution. The contour of the electrostatic potential is shown (green dashed curves).  The leads $S$ and $D$ are also labelled.}
\end{figure}

\section{The electric fields and the FIR spectra}

\subsection{Three typical in-plane electric fields}

In Fig. \ref{fig:EF}, three typical in-plane electric fields are applied respectively to a TGQD with the size $N_s$ = 8, where $N_s$ is the number of carbon atoms in each side of the dot.
Each electric field is generated by two gate electrodes with opposite electrostatic potentials $\pm U$.
In the following, the symmetry characteristics of the three electric fields are analyzed simply.
In Fig. \ref{fig:EF}(a), the triangular electric field $EF1$ possesses the same $C3$ rotation symmetry as the quantum dot.
In Fig. \ref{fig:EF}(b), the uniform electric field $EF2$ damages the $C3$ symmetry of the system even though it is homogeneous.
In Fig. \ref{fig:EF}(c), the random electric field $EF3$ presents randomly an imaginary potential distribution, which simulates an electric field with irregular gate electrodes.
In contrast, $EF1$ does not change the symmetry of the system while $EF3$ causes the most serious damage to the symmetry of the system.

\subsection{The FIR spectra and the selection rules}

\begin{figure*}[htp]
\centering
\includegraphics[height=11cm]{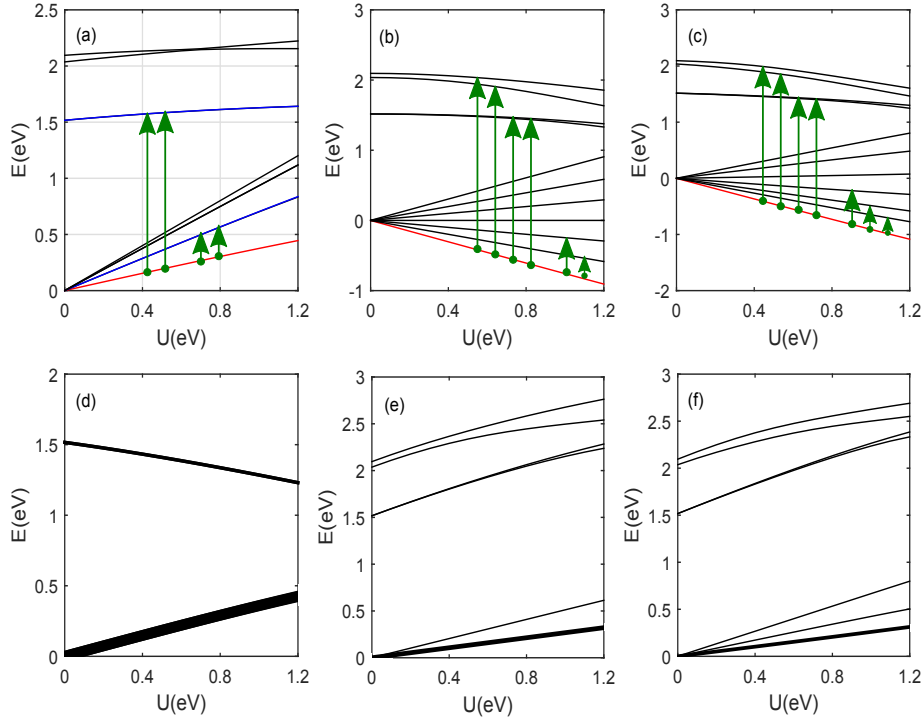}\caption{\label{fig:Fig2}
The energy spectra and the FIR spectra of a TGQD ($N_{s}$ = 8). (a) The energy spectra with $EF1$ where the blue lines correspond to the double degenerate levels. (b) The energy spectra with $EF2$. (c) The energy spectra with $EF3$. (d) The FIR spectra with $EF1$. (e) The FIR spectra with $EF2$. (f) The FIR spectra with $EF3$. In (a), (b) and (c), the red lines correspond to the ground-state levels and the green arrows indicate the selection rules. In (d), (e) and (f), the line width is roughly proportional to the peak intensity.}
\end{figure*}

\begin{figure*}[htp]
\centering
\includegraphics[height=9cm]{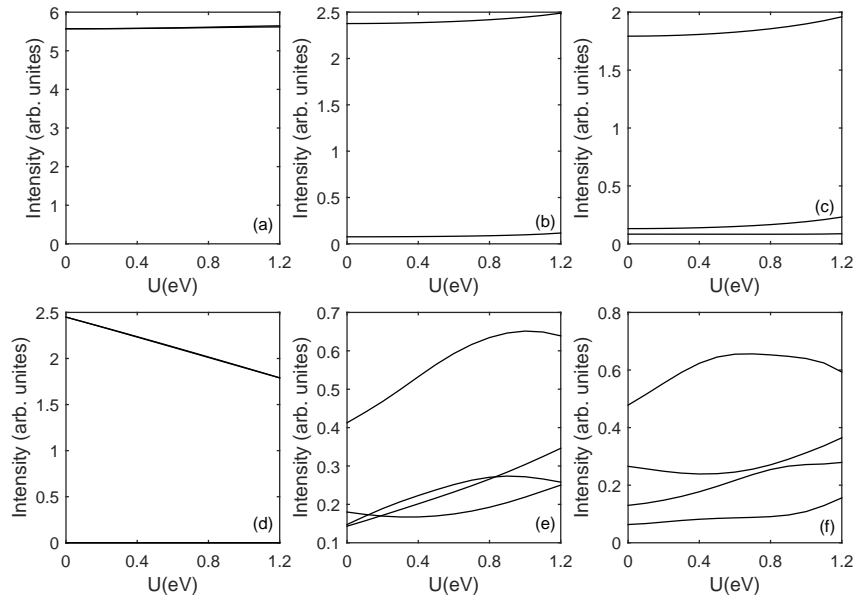}\caption{\label{fig:Fig3}
The intensities of the absorption peaks for the intraband and interband transition in a TGQD ($N_{s}$ = 8). (a) The intraband transition with $EF1$. (b) The intraband transition with $EF2$. (c) The intraband transition with $EF3$. (d) The interband transition with $EF1$. (e) The interband transition with $EF2$. (f) The interband transition with $EF3$.}
\end{figure*}

Fig. \ref{fig:Fig2} shows the energy spectra and the calculated FIR spectra of a single-electron TGQD ($N_{s}$ = 8).
The energy spectra are shown in Fig. \ref{fig:Fig2}(a-c) and the more details of the energy spectra can be seen elsewhere\cite{dong2014electronic}.
According to the ground-state level and the chemical potential of the leads, one can guarantee that there is only one electron in the dot.
The calculated FIR spectra are shown in Fig. \ref{fig:Fig2}(d-f) and the corresponding optical selection rules are marked on the energy spectra.
In the FIR spectra, we consider only the optical transitions where the excited states are the zero-energy band and the four lowest states of the non-zero band.
Moreover, we have included only the transitions which have a peak intensity of more than 1\% of the maximum value.
Also, we have plotted the intensities of the absorption peaks for the intraband transitions in Fig. \ref{fig:Fig3}(a-c) and those for the interband transitions in Fig. \ref{fig:Fig3}(e-f).
Fig. \ref{fig:Fig3} makes it easier to compare the difference between the intraband transitions and the interband transitions.
As a general feature of the calculated spectra shown in Fig. \ref{fig:Fig2}(d-f), one can see that each spectrum has two branches as a major component, where the higher one comes from the interband transitions and the lower from the intraband transitions.
According to the energy spectra, the specific selection rules can be identified easily.
These forbidden transitions are attributed to the $C3$ rotational symmetry of the dot.

The effects of three electric fields on the FIR spectra are compared in the following.
The electric field $EF1$ possesses a $C3$ rotational symmetry and thus the $C3$ symmetry of the system is not damaged.
Fig. \ref{fig:Fig2}(d) shows the FIR spectra of the dot subjected to the triangular electric field $EF1$.
In the absorption spectrum, there are only two absorption peaks.
One peak comes from the intraband transition and the other comes from the interband transition.
The selection rule is marked in Fig. \ref{fig:Fig2}(a).
Fig. \ref{fig:Fig3}(a, d) shows that the intensity of the interband peak is less than half the intensity of the intraband peak.
It should be noted that the excited level of the intraband transition is double degenerate and the two degenerate states contribute the same peak intensity.
Therefore, the peak intensity should be multiplied by two if the data are measured experimentally.
This kind of degeneracy can also be seen in the interband transition.
Later it will be shown that the $x$ or $y$ polarization breaks the balance of the peak intensities due to the degeneracy.

Fig. \ref{fig:Fig2}(e) shows the FIR spectra of the dot subjected to the uniform electric field $EF2$.
There appear two absorption peaks in the intraband transition and Fig. \ref{fig:Fig3}(b) shows the intensity of the second intraband peak is about 5\% of the intensity of the strongest intraband peak.
The transitions from the ground state to the non-zero band are all allowed.
The selection rule is marked in Fig. \ref{fig:Fig2}(b).
Fig. \ref{fig:Fig3}(b,e) shows that the intensity of the strongest interband peak is about 25\% of the intensity of the strongest intraband peak.
In contrast to the situation with $EF1$, both the intensity of the strongest peak and the number of the forbidden transitions decrease significantly.
The phenomenon suggests that the $C3$ symmetry of the system has been damaged to a certain extent.

Fig. \ref{fig:Fig2}(f) shows the FIR spectra of the dot subjected to the random electric field $EF3$.
There appear three intraband absorption peaks in the spectra.
Fig. \ref{fig:Fig3}(c) shows that the intensity of the second intraband peak is about 10\% of the intensity of the strongest intraband peak and the intensity of the third intraband peak is about 5\% of the intensity of the strongest intraband peak.
The transitions from the ground state to the non-zero band are all allowed.
The selection rule is marked in Fig. \ref{fig:Fig2}(c).
Fig. \ref{fig:Fig3}(c,f) shows that the intensity of the strongest interband peak is about 30\% of the intensity of the strongest intraband peak.
In contrast to the situation with $EF2$, the intensity of the strongest peak and the number of the forbidden transitions decreases further.
The phenomenon suggests that the disorder of the random electric field has damaged further the $C3$ symmetry of the system.

From the electric field $EF1$ to $EF2$ and then to $EF3$, the $C3$ symmetry of the system is damaged gradually.
Therefore, both the intensity of the strongest peak and the number of the forbidden transitions decrease gradually.
These phenomenons can also be explained in view of the wave function.
As the symmetry is damaged, the eigenstates are recombined and the wavefunction component that allows the transition are dispersed, which leads to more absorption peaks.
The  intensities of the intraband peaks are almost constant with $U$ while the intensities of the interband peaks change drastically with $U$.
The reason is that the eigenstates of the zero-energy band are almost constant with $U$ while the eigenstates of the nonzero-energy band are mixed continuously with $U$\cite{dong2014electronic}.

\subsection{The effect of polarization on the FIR spectra}

\begin{figure}[htp]
\centering
\includegraphics[height=17cm]{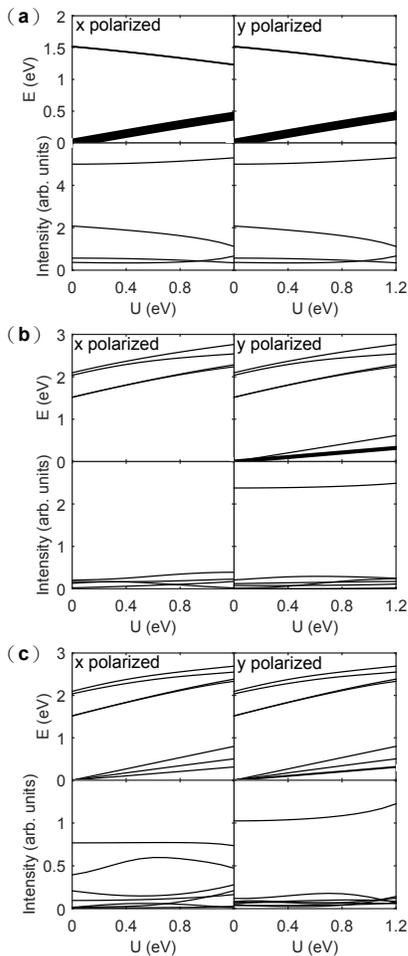}\caption{\label{fig:Fig4}
The $x$- and $y$-polarized FIR spectra of a TGQD ($N_{s}$ = 8) (a) subjected to $EF1$, (b) subjected to $EF2$ and (c) subjected to $EF3$. In each panel, the left side corresponds to the $x$ polarization, the right side to the $y$ polarization, the upper side to the peak energy and the lower side to the peak intensity. The width of each line for the peak energy is roughly proportional to the corresponding peak intensity.}
\end{figure}

In the following, we investigate the effect of $x$ and $y$ polarization on the FIR spectra by comparing the polarized spectra with the unpolarized spectra.
Fig. \ref{fig:Fig4}(a) shows the $x$- and $y$-polarized FIR spectra of the dot subjected to the electric field $EF1$.
Compared with the unpolarized spectra, the peak energies of the $x$- and $y$-polarized spectra do not change while the peak intensities change significantly.
The balances of the peak intensities due to the degeneracy are broken since the polarization reduces some relevant transition matrix element.
Although the system is asymmetric in the $x$ and $y$ directions, the effects of the $x$ and $y$ polarization on the spectra are similar.
This coincidence may be related to the specific distribution of the wave function.
Fig. \ref{fig:Fig4}(b) shows the $x$- and $y$-polarized FIR spectra of the dot subjected to the electric field $EF2$.
A remarkable phenomenon is that all the intraband transitions are forbidden in the $x$-polarized spectrum.
In the $y$-polarized spectrum, the peak energies are the same as the unpolarized spectra as shown in Fig. \ref{fig:Fig2}(e).
The peak intensities of the unpolarized spectra are allocated unequally to the $x$- and $y$-polarized spectra, which is consistent with Eq. (\ref{eqn:2}) and Eq. (\ref{eqn:3}).
Fig. \ref{fig:Fig4}(c) shows the $x$- and $y$-polarized FIR spectra of the dot subjected to the electric field $EF3$.
The peak energies in the $x$- and $y$-polarized spectra are the same as the unpolarized spectra as shown in Fig. \ref{fig:Fig2}(f).
In other words, the selection rules are not changed by the $x$ and $y$ polarization.
The peak intensities of the unpolarized spectra are allocated roughly equally to the $x$- and $y$-polarized spectra.
This fact shows that the polarization can not generate any new forbidden transition.
This phenomenon implies that the random electric field causes more damage to the $C3$ symmetry than the electric fields $EF2$.
By comparing the polarized spectra with the unpolarized spectra, it can be seen that the polarization causes the decrease of the peak intensities and even the new forbidden transitions.
Moreover, the effects of the polarization are related closely to the symmetry of the electric fields.

\section{Summary}

In this paper, we have investigated the effects of three typical in-plane electric fields on the FIR spectra of a single-electron triangular zigzag graphene quantum dot.
Our result shows that there exist the remarkable selection rules in the FIR spectra due to the $C3$ symmetry of the dot.
When the electric field possesses also the $C3$ symmetry, there are only two absorption peaks.
As the $C3$ symmetry of the system is damaged by the electric fields, both the intensity of the strongest peak and the number of the forbidden transitions decrease gradually.
The intensities of the intraband peaks are almost constant with $U$ while the intensities of the interband peaks change drastically.
The polarization causes the decrease of the peak intensities and even the new forbidden transitions.
Our findings may help to probe the electronic structure of GQDs by FIR spectroscopy and may be useful for the application of GQDs to electronic and optoelectronic devices.

\section*{Acknowledgements}
This work is supported by the National Natural Science Foundation of China (Grant Nos. 11604183 and 11674197), and a Project of Shandong Province Higher Educational Science and Technology Program (Grant No. J16LJ09).


\section*{References}

\begin{thebibliography}{40}
\expandafter\ifx\csname natexlab\endcsname\relax\def\natexlab#1{#1}\fi
\expandafter\ifx\csname bibnamefont\endcsname\relax
  \def\bibnamefont#1{#1}\fi
\expandafter\ifx\csname bibfnamefont\endcsname\relax
  \def\bibfnamefont#1{#1}\fi
\expandafter\ifx\csname citenamefont\endcsname\relax
  \def\citenamefont#1{#1}\fi
\expandafter\ifx\csname url\endcsname\relax
  \def\url#1{\texttt{#1}}\fi
\expandafter\ifx\csname urlprefix\endcsname\relax\def\urlprefix{URL }\fi
\providecommand{\bibinfo}[2]{#2}
\providecommand{\eprint}[2][]{\url{#2}}

\bibitem[{\citenamefont{Novoselov et~al.}(2004)\citenamefont{Novoselov, Geim,
  Morozov, Jiang, Zhang, Dubonos, Grigorieva, and Firsov}}]{Novoselov04}
\bibinfo{author}{\bibfnamefont{K.~S.} \bibnamefont{Novoselov}},
  \bibinfo{author}{\bibfnamefont{A.~K.} \bibnamefont{Geim}},
  \bibinfo{author}{\bibfnamefont{S.~V.} \bibnamefont{Morozov}},
  \bibinfo{author}{\bibfnamefont{D.}~\bibnamefont{Jiang}},
  \bibinfo{author}{\bibfnamefont{Y.}~\bibnamefont{Zhang}},
  \bibinfo{author}{\bibfnamefont{S.~V.} \bibnamefont{Dubonos}},
  \bibinfo{author}{\bibfnamefont{I.~V.} \bibnamefont{Grigorieva}},
  \bibnamefont{and} \bibinfo{author}{\bibfnamefont{A.~A.}
  \bibnamefont{Firsov}}, \bibinfo{journal}{Science}
  \textbf{\bibinfo{volume}{306}}, \bibinfo{pages}{666} (\bibinfo{year}{2004}).

\bibitem[{\citenamefont{Katsnelson et~al.}(2006)\citenamefont{Katsnelson,
  Novoselov, and Geim}}]{katsnelson2006chiral}
\bibinfo{author}{\bibfnamefont{M.~I.} \bibnamefont{Katsnelson}},
  \bibinfo{author}{\bibfnamefont{K.~S.} \bibnamefont{Novoselov}},
  \bibnamefont{and} \bibinfo{author}{\bibfnamefont{A.~K.} \bibnamefont{Geim}},
  \bibinfo{journal}{Nature physics} \textbf{\bibinfo{volume}{2}},
  \bibinfo{pages}{620} (\bibinfo{year}{2006}).

\bibitem[{\citenamefont{Bolotin et~al.}(2008)\citenamefont{Bolotin, Sikes,
  Jiang, Klima, Fudenberg, Kim, and Stormer}}]{bolotin2008ultrahigh}
\bibinfo{author}{\bibfnamefont{K.~I.} \bibnamefont{Bolotin}},
  \bibinfo{author}{\bibfnamefont{K.~J.} \bibnamefont{Sikes}},
  \bibinfo{author}{\bibfnamefont{Z.}~\bibnamefont{Jiang}},
  \bibinfo{author}{\bibfnamefont{M.}~\bibnamefont{Klima}},
  \bibinfo{author}{\bibfnamefont{G.}~\bibnamefont{Fudenberg}},
  \bibinfo{author}{\bibfnamefont{J.~H.~P.} \bibnamefont{Kim}},
  \bibnamefont{and} \bibinfo{author}{\bibfnamefont{H.~L.}
  \bibnamefont{Stormer}}, \bibinfo{journal}{Solid State Communications}
  \textbf{\bibinfo{volume}{146}}, \bibinfo{pages}{351} (\bibinfo{year}{2008}).

\bibitem[{\citenamefont{Balandin et~al.}(2008)\citenamefont{Balandin, Ghosh,
  Bao, Calizo, Teweldebrhan, Miao, and Lau}}]{Balandin2008}
\bibinfo{author}{\bibfnamefont{A.~A.} \bibnamefont{Balandin}},
  \bibinfo{author}{\bibfnamefont{S.}~\bibnamefont{Ghosh}},
  \bibinfo{author}{\bibfnamefont{W.}~\bibnamefont{Bao}},
  \bibinfo{author}{\bibfnamefont{I.}~\bibnamefont{Calizo}},
  \bibinfo{author}{\bibfnamefont{D.}~\bibnamefont{Teweldebrhan}},
  \bibinfo{author}{\bibfnamefont{F.}~\bibnamefont{Miao}}, \bibnamefont{and}
  \bibinfo{author}{\bibfnamefont{C.~N.} \bibnamefont{Lau}},
  \bibinfo{journal}{Nano letters} \textbf{\bibinfo{volume}{8}},
  \bibinfo{pages}{902} (\bibinfo{year}{2008}).

\bibitem[{\citenamefont{Lee et~al.}(2008)\citenamefont{Lee, Wei, Kysar, and
  Hone}}]{lee2008measurement}
\bibinfo{author}{\bibfnamefont{C.}~\bibnamefont{Lee}},
  \bibinfo{author}{\bibfnamefont{X.}~\bibnamefont{Wei}},
  \bibinfo{author}{\bibfnamefont{J.~W.} \bibnamefont{Kysar}}, \bibnamefont{and}
  \bibinfo{author}{\bibfnamefont{J.}~\bibnamefont{Hone}},
  \bibinfo{journal}{science} \textbf{\bibinfo{volume}{321}},
  \bibinfo{pages}{385} (\bibinfo{year}{2008}).

\bibitem[{\citenamefont{Nair et~al.}(2008)\citenamefont{Nair, Blake,
  Grigorenko, Novoselov, Booth, Stauber, Peres, and Geim}}]{nair2008fine}
\bibinfo{author}{\bibfnamefont{R.~R.} \bibnamefont{Nair}},
  \bibinfo{author}{\bibfnamefont{P.}~\bibnamefont{Blake}},
  \bibinfo{author}{\bibfnamefont{A.~N.} \bibnamefont{Grigorenko}},
  \bibinfo{author}{\bibfnamefont{K.~S.} \bibnamefont{Novoselov}},
  \bibinfo{author}{\bibfnamefont{T.~J.} \bibnamefont{Booth}},
  \bibinfo{author}{\bibfnamefont{T.}~\bibnamefont{Stauber}},
  \bibinfo{author}{\bibfnamefont{N.~M.} \bibnamefont{Peres}}, \bibnamefont{and}
  \bibinfo{author}{\bibfnamefont{A.~K.} \bibnamefont{Geim}},
  \bibinfo{journal}{Science} \textbf{\bibinfo{volume}{320}},
  \bibinfo{pages}{1308} (\bibinfo{year}{2008}).

\bibitem[{\citenamefont{Novoselov et~al.}(2012)\citenamefont{Novoselov, Fal'ko,
  Colombo, Gellert, Schwab, and Kim}}]{Novoselov12}
\bibinfo{author}{\bibfnamefont{K.~S.} \bibnamefont{Novoselov}},
  \bibinfo{author}{\bibfnamefont{V.~I.} \bibnamefont{Fal'ko}},
  \bibinfo{author}{\bibfnamefont{L.}~\bibnamefont{Colombo}},
  \bibinfo{author}{\bibfnamefont{P.~R.} \bibnamefont{Gellert}},
  \bibinfo{author}{\bibfnamefont{M.~G.} \bibnamefont{Schwab}},
  \bibnamefont{and} \bibinfo{author}{\bibfnamefont{K.}~\bibnamefont{Kim}},
  \bibinfo{journal}{Nature} \textbf{\bibinfo{volume}{490}},
  \bibinfo{pages}{192} (\bibinfo{year}{2012}).

\bibitem[{\citenamefont{Xia et~al.}(2014)\citenamefont{Xia, Wang, Xiao, Dubey,
  and Ramasubramaniam}}]{Xia14}
\bibinfo{author}{\bibfnamefont{F.}~\bibnamefont{Xia}},
  \bibinfo{author}{\bibfnamefont{H.}~\bibnamefont{Wang}},
  \bibinfo{author}{\bibfnamefont{D.}~\bibnamefont{Xiao}},
  \bibinfo{author}{\bibfnamefont{M.}~\bibnamefont{Dubey}}, \bibnamefont{and}
  \bibinfo{author}{\bibfnamefont{A.}~\bibnamefont{Ramasubramaniam}},
  \bibinfo{journal}{Nat. Photonics} \textbf{\bibinfo{volume}{8}},
  \bibinfo{pages}{899} (\bibinfo{year}{2014}).

\bibitem[{\citenamefont{Neto et~al.}(2009)\citenamefont{Neto, Guinea, Peres,
  Novoselov, and Geim}}]{neto2009electronic}
\bibinfo{author}{\bibfnamefont{A.~C.} \bibnamefont{Neto}},
  \bibinfo{author}{\bibfnamefont{F.}~\bibnamefont{Guinea}},
  \bibinfo{author}{\bibfnamefont{N.~M.} \bibnamefont{Peres}},
  \bibinfo{author}{\bibfnamefont{K.~S.} \bibnamefont{Novoselov}},
  \bibnamefont{and} \bibinfo{author}{\bibfnamefont{A.~K.} \bibnamefont{Geim}},
  \bibinfo{journal}{Reviews of modern physics} \textbf{\bibinfo{volume}{81}},
  \bibinfo{pages}{109} (\bibinfo{year}{2009}).

\bibitem[{\citenamefont{Sarma et~al.}(2011)\citenamefont{Sarma, Adam, Hwang,
  and Rossi}}]{sarma2011electronic}
\bibinfo{author}{\bibfnamefont{S.~D.} \bibnamefont{Sarma}},
  \bibinfo{author}{\bibfnamefont{S.}~\bibnamefont{Adam}},
  \bibinfo{author}{\bibfnamefont{E.}~\bibnamefont{Hwang}}, \bibnamefont{and}
  \bibinfo{author}{\bibfnamefont{E.}~\bibnamefont{Rossi}},
  \bibinfo{journal}{Reviews of Modern Physics} \textbf{\bibinfo{volume}{83}},
  \bibinfo{pages}{407} (\bibinfo{year}{2011}).

\bibitem[{\citenamefont{Kotov et~al.}(2012)\citenamefont{Kotov, Uchoa, Pereira,
  Guinea, and Neto}}]{kotov2012electron}
\bibinfo{author}{\bibfnamefont{V.~N.} \bibnamefont{Kotov}},
  \bibinfo{author}{\bibfnamefont{B.}~\bibnamefont{Uchoa}},
  \bibinfo{author}{\bibfnamefont{V.~M.} \bibnamefont{Pereira}},
  \bibinfo{author}{\bibfnamefont{F.}~\bibnamefont{Guinea}}, \bibnamefont{and}
  \bibinfo{author}{\bibfnamefont{A.~C.} \bibnamefont{Neto}},
  \bibinfo{journal}{Reviews of Modern Physics} \textbf{\bibinfo{volume}{84}},
  \bibinfo{pages}{1067} (\bibinfo{year}{2012}).

\bibitem[{\citenamefont{G{\"u}{\c{c}}l{\"u}
  et~al.}(2014)\citenamefont{G{\"u}{\c{c}}l{\"u}, Potasz, Korkusinski, and
  Hawrylak}}]{gucclu2014graphene}
\bibinfo{author}{\bibfnamefont{A.~D.} \bibnamefont{G{\"u}{\c{c}}l{\"u}}},
  \bibinfo{author}{\bibfnamefont{P.}~\bibnamefont{Potasz}},
  \bibinfo{author}{\bibfnamefont{M.}~\bibnamefont{Korkusinski}},
  \bibnamefont{and} \bibinfo{author}{\bibfnamefont{P.}~\bibnamefont{Hawrylak}},
  \emph{\bibinfo{title}{Graphene quantum dots}} (\bibinfo{publisher}{Springer},
  \bibinfo{year}{2014}).

\bibitem[{\citenamefont{Liu et~al.}(2013)\citenamefont{Liu, Feng, Yan, Chen,
  and Xue}}]{liu2013superior}
\bibinfo{author}{\bibfnamefont{W.-W.} \bibnamefont{Liu}},
  \bibinfo{author}{\bibfnamefont{Y.-Q.} \bibnamefont{Feng}},
  \bibinfo{author}{\bibfnamefont{X.-B.} \bibnamefont{Yan}},
  \bibinfo{author}{\bibfnamefont{J.-T.} \bibnamefont{Chen}}, \bibnamefont{and}
  \bibinfo{author}{\bibfnamefont{Q.-J.} \bibnamefont{Xue}},
  \bibinfo{journal}{Advanced Functional Materials}
  \textbf{\bibinfo{volume}{23}}, \bibinfo{pages}{4111} (\bibinfo{year}{2013}).

\bibitem[{\citenamefont{Joo et~al.}(2014)\citenamefont{Joo, Kim, Kang, Kim,
  Choi, and Hwang}}]{joo2014graphene}
\bibinfo{author}{\bibfnamefont{S.~S.} \bibnamefont{Joo}},
  \bibinfo{author}{\bibfnamefont{J.}~\bibnamefont{Kim}},
  \bibinfo{author}{\bibfnamefont{S.~S.} \bibnamefont{Kang}},
  \bibinfo{author}{\bibfnamefont{S.}~\bibnamefont{Kim}},
  \bibinfo{author}{\bibfnamefont{S.-H.} \bibnamefont{Choi}}, \bibnamefont{and}
  \bibinfo{author}{\bibfnamefont{S.~W.} \bibnamefont{Hwang}},
  \bibinfo{journal}{Nanotechnology} \textbf{\bibinfo{volume}{25}},
  \bibinfo{pages}{255203} (\bibinfo{year}{2014}).

\bibitem[{\citenamefont{Kim et~al.}(2014)\citenamefont{Kim, Hwang, Kim, Shin,
  Kang, Kim, Jang, Kim, Lee, Choi et~al.}}]{kim2014high}
\bibinfo{author}{\bibfnamefont{C.~O.} \bibnamefont{Kim}},
  \bibinfo{author}{\bibfnamefont{S.~W.} \bibnamefont{Hwang}},
  \bibinfo{author}{\bibfnamefont{S.}~\bibnamefont{Kim}},
  \bibinfo{author}{\bibfnamefont{D.~H.} \bibnamefont{Shin}},
  \bibinfo{author}{\bibfnamefont{S.~S.} \bibnamefont{Kang}},
  \bibinfo{author}{\bibfnamefont{J.~M.} \bibnamefont{Kim}},
  \bibinfo{author}{\bibfnamefont{C.~W.} \bibnamefont{Jang}},
  \bibinfo{author}{\bibfnamefont{J.~H.} \bibnamefont{Kim}},
  \bibinfo{author}{\bibfnamefont{K.~W.} \bibnamefont{Lee}},
  \bibinfo{author}{\bibfnamefont{S.-H.} \bibnamefont{Choi}},
  \bibnamefont{et~al.}, \bibinfo{journal}{Scientific reports}
  \textbf{\bibinfo{volume}{4}}, \bibinfo{pages}{5603} (\bibinfo{year}{2014}).

\bibitem[{\citenamefont{Konstantatos et~al.}(2012)\citenamefont{Konstantatos,
  Badioli, Gaudreau, Osmond, Bernechea, De~Arquer, Gatti, and
  Koppens}}]{konstantatos2012hybrid}
\bibinfo{author}{\bibfnamefont{G.}~\bibnamefont{Konstantatos}},
  \bibinfo{author}{\bibfnamefont{M.}~\bibnamefont{Badioli}},
  \bibinfo{author}{\bibfnamefont{L.}~\bibnamefont{Gaudreau}},
  \bibinfo{author}{\bibfnamefont{J.}~\bibnamefont{Osmond}},
  \bibinfo{author}{\bibfnamefont{M.}~\bibnamefont{Bernechea}},
  \bibinfo{author}{\bibfnamefont{F.~P.~G.} \bibnamefont{De~Arquer}},
  \bibinfo{author}{\bibfnamefont{F.}~\bibnamefont{Gatti}}, \bibnamefont{and}
  \bibinfo{author}{\bibfnamefont{F.~H.} \bibnamefont{Koppens}},
  \bibinfo{journal}{Nature nanotechnology} \textbf{\bibinfo{volume}{7}},
  \bibinfo{pages}{363} (\bibinfo{year}{2012}).

\bibitem[{\citenamefont{Bacon et~al.}(2014)\citenamefont{Bacon, Bradley, and
  Nann}}]{bacon2014graphene}
\bibinfo{author}{\bibfnamefont{M.}~\bibnamefont{Bacon}},
  \bibinfo{author}{\bibfnamefont{S.~J.} \bibnamefont{Bradley}},
  \bibnamefont{and} \bibinfo{author}{\bibfnamefont{T.}~\bibnamefont{Nann}},
  \bibinfo{journal}{Particle \& Particle Systems Characterization}
  \textbf{\bibinfo{volume}{31}}, \bibinfo{pages}{415} (\bibinfo{year}{2014}).

\bibitem[{\citenamefont{G\"{u}cl\"{u} et~al.}(2009)\citenamefont{G\"{u}cl\"{u},
  Potasz, Voznyy, Korkusinski, and Hawrylak}}]{Gu09}
\bibinfo{author}{\bibfnamefont{A.~D.} \bibnamefont{G\"{u}cl\"{u}}},
  \bibinfo{author}{\bibfnamefont{P.}~\bibnamefont{Potasz}},
  \bibinfo{author}{\bibfnamefont{O.}~\bibnamefont{Voznyy}},
  \bibinfo{author}{\bibfnamefont{M.}~\bibnamefont{Korkusinski}},
  \bibnamefont{and} \bibinfo{author}{\bibfnamefont{P.}~\bibnamefont{Hawrylak}},
  \bibinfo{journal}{Phys. Rev. Lett.} \textbf{\bibinfo{volume}{103}},
  \bibinfo{pages}{246805} (\bibinfo{year}{2009}).

\bibitem[{\citenamefont{Abergel
  et~al.}(2010{\natexlab{a}})\citenamefont{Abergel, Apalkov, Berashevich,
  Ziegler, and Chakraborty}}]{abergel2010properties}
\bibinfo{author}{\bibfnamefont{D.}~\bibnamefont{Abergel}},
  \bibinfo{author}{\bibfnamefont{V.}~\bibnamefont{Apalkov}},
  \bibinfo{author}{\bibfnamefont{J.}~\bibnamefont{Berashevich}},
  \bibinfo{author}{\bibfnamefont{K.}~\bibnamefont{Ziegler}}, \bibnamefont{and}
  \bibinfo{author}{\bibfnamefont{T.}~\bibnamefont{Chakraborty}},
  \bibinfo{journal}{Advances in Physics} \textbf{\bibinfo{volume}{59}},
  \bibinfo{pages}{261} (\bibinfo{year}{2010}{\natexlab{a}}).

\bibitem[{\citenamefont{Potasz et~al.}(2012)\citenamefont{Potasz,
  G\"{u}cl\"{u}, A.W\'{o}js, and Hawrylak}}]{Potasz12}
\bibinfo{author}{\bibfnamefont{P.}~\bibnamefont{Potasz}},
  \bibinfo{author}{\bibfnamefont{A.~D.} \bibnamefont{G\"{u}cl\"{u}}},
  \bibinfo{author}{\bibnamefont{A.W\'{o}js}}, \bibnamefont{and}
  \bibinfo{author}{\bibfnamefont{P.}~\bibnamefont{Hawrylak}},
  \bibinfo{journal}{Phys. Rev. B} \textbf{\bibinfo{volume}{85}},
  \bibinfo{pages}{075431} (\bibinfo{year}{2012}).

\bibitem[{\citenamefont{Potasz et~al.}(2010)\citenamefont{Potasz,
  G{\"u}{\c{c}}l{\"u}, and Hawrylak}}]{potasz2010zero}
\bibinfo{author}{\bibfnamefont{P.}~\bibnamefont{Potasz}},
  \bibinfo{author}{\bibfnamefont{A.~D.} \bibnamefont{G{\"u}{\c{c}}l{\"u}}},
  \bibnamefont{and} \bibinfo{author}{\bibfnamefont{P.}~\bibnamefont{Hawrylak}},
  \bibinfo{journal}{Physical review. B} \textbf{\bibinfo{volume}{81}},
  \bibinfo{pages}{033403} (\bibinfo{year}{2010}).

\bibitem[{\citenamefont{Liang et~al.}(2014)\citenamefont{Liang, Jiang, Lv,
  Zhang, and Li}}]{JZT2014}
\bibinfo{author}{\bibfnamefont{F.~X.} \bibnamefont{Liang}},
  \bibinfo{author}{\bibfnamefont{Z.~T.} \bibnamefont{Jiang}},
  \bibinfo{author}{\bibfnamefont{Z.~T.} \bibnamefont{Lv}},
  \bibinfo{author}{\bibfnamefont{H.~Y.} \bibnamefont{Zhang}}, \bibnamefont{and}
  \bibinfo{author}{\bibfnamefont{S.}~\bibnamefont{Li}},
  \bibinfo{journal}{Journal of Applied Physics} \textbf{\bibinfo{volume}{116}},
  \bibinfo{pages}{123706} (\bibinfo{year}{2014}).

\bibitem[{\citenamefont{Sun et~al.}(2017)\citenamefont{Sun, Zheng, Pan, Chen,
  Zhang, Fu, Zhang, Tang, and Du}}]{sun2017magnetism}
\bibinfo{author}{\bibfnamefont{Y.}~\bibnamefont{Sun}},
  \bibinfo{author}{\bibfnamefont{Y.}~\bibnamefont{Zheng}},
  \bibinfo{author}{\bibfnamefont{H.}~\bibnamefont{Pan}},
  \bibinfo{author}{\bibfnamefont{J.}~\bibnamefont{Chen}},
  \bibinfo{author}{\bibfnamefont{W.}~\bibnamefont{Zhang}},
  \bibinfo{author}{\bibfnamefont{L.}~\bibnamefont{Fu}},
  \bibinfo{author}{\bibfnamefont{K.}~\bibnamefont{Zhang}},
  \bibinfo{author}{\bibfnamefont{N.}~\bibnamefont{Tang}}, \bibnamefont{and}
  \bibinfo{author}{\bibfnamefont{Y.}~\bibnamefont{Du}}, \bibinfo{journal}{npj
  Quantum Materials} \textbf{\bibinfo{volume}{2}}, \bibinfo{pages}{5}
  (\bibinfo{year}{2017}).

\bibitem[{\citenamefont{Basak and Shukla}(2016)}]{basak2016optical}
\bibinfo{author}{\bibfnamefont{T.}~\bibnamefont{Basak}} \bibnamefont{and}
  \bibinfo{author}{\bibfnamefont{A.}~\bibnamefont{Shukla}},
  \bibinfo{journal}{Physical Review B} \textbf{\bibinfo{volume}{93}},
  \bibinfo{pages}{235432} (\bibinfo{year}{2016}).

\bibitem[{\citenamefont{Hawrylak et~al.}(2016)\citenamefont{Hawrylak, Peeters,
  and Ensslin}}]{hawrylak2016carbononics}
\bibinfo{author}{\bibfnamefont{P.}~\bibnamefont{Hawrylak}},
  \bibinfo{author}{\bibfnamefont{F.}~\bibnamefont{Peeters}}, \bibnamefont{and}
  \bibinfo{author}{\bibfnamefont{K.}~\bibnamefont{Ensslin}},
  \bibinfo{journal}{physica status solidi (RRL)-Rapid Research Letters}
  \textbf{\bibinfo{volume}{10}}, \bibinfo{pages}{11} (\bibinfo{year}{2016}).

\bibitem[{\citenamefont{Farghadan and Saffarzadeh}(2014)}]{farghadan2014Iran}
\bibinfo{author}{\bibfnamefont{R.}~\bibnamefont{Farghadan}} \bibnamefont{and}
  \bibinfo{author}{\bibfnamefont{A.}~\bibnamefont{Saffarzadeh}},
  \bibinfo{journal}{Journal of Applied Physics} \textbf{\bibinfo{volume}{115}},
  \bibinfo{pages}{174310} (\bibinfo{year}{2014}).

\bibitem[{\citenamefont{Chen et~al.}(2010)\citenamefont{Chen, Chang, and
  Lin}}]{Chen10}
\bibinfo{author}{\bibfnamefont{R.~B.} \bibnamefont{Chen}},
  \bibinfo{author}{\bibfnamefont{C.~P.} \bibnamefont{Chang}}, \bibnamefont{and}
  \bibinfo{author}{\bibfnamefont{M.~F.} \bibnamefont{Lin}},
  \bibinfo{journal}{Physica E} \textbf{\bibinfo{volume}{42}},
  \bibinfo{pages}{2812} (\bibinfo{year}{2010}).

\bibitem[{\citenamefont{Ma and Li}(2012{\natexlab{a}})}]{Ma12PRB}
\bibinfo{author}{\bibfnamefont{W.~L.} \bibnamefont{Ma}} \bibnamefont{and}
  \bibinfo{author}{\bibfnamefont{S.~S.} \bibnamefont{Li}},
  \bibinfo{journal}{Phys. Rev. B} \textbf{\bibinfo{volume}{86}},
  \bibinfo{pages}{045449} (\bibinfo{year}{2012}{\natexlab{a}}).

\bibitem[{\citenamefont{Dong}(2013)}]{Dong13}
\bibinfo{author}{\bibfnamefont{Q.~R.} \bibnamefont{Dong}}, \bibinfo{journal}{J.
  Appl. Phys.} \textbf{\bibinfo{volume}{113}}, \bibinfo{pages}{234304}
  (\bibinfo{year}{2013}).

\bibitem[{\citenamefont{G\"{u}cl\"{u} and Hawrylak}(2013)}]{Gu13}
\bibinfo{author}{\bibfnamefont{A.~D.} \bibnamefont{G\"{u}cl\"{u}}}
  \bibnamefont{and} \bibinfo{author}{\bibfnamefont{P.}~\bibnamefont{Hawrylak}},
  \bibinfo{journal}{Phys. Rev. B} \textbf{\bibinfo{volume}{87}},
  \bibinfo{pages}{035425} (\bibinfo{year}{2013}).

\bibitem[{\citenamefont{Sza{\l}owski}(2015)}]{szalowski2015Poland}
\bibinfo{author}{\bibfnamefont{K.}~\bibnamefont{Sza{\l}owski}},
  \bibinfo{journal}{Journal of Magnetism and Magnetic Materials}
  \textbf{\bibinfo{volume}{382}}, \bibinfo{pages}{318} (\bibinfo{year}{2015}).

\bibitem[{\citenamefont{Li et~al.}(2017)\citenamefont{Li, Zarenia, Xu, Dong,
  and Peeters}}]{Peeters2017exciton}
\bibinfo{author}{\bibfnamefont{L.~L.} \bibnamefont{Li}},
  \bibinfo{author}{\bibfnamefont{M.}~\bibnamefont{Zarenia}},
  \bibinfo{author}{\bibfnamefont{W.}~\bibnamefont{Xu}},
  \bibinfo{author}{\bibfnamefont{H.~M.} \bibnamefont{Dong}}, \bibnamefont{and}
  \bibinfo{author}{\bibfnamefont{F.~M.} \bibnamefont{Peeters}},
  \bibinfo{journal}{Physical Review B} \textbf{\bibinfo{volume}{95}},
  \bibinfo{pages}{045409} (\bibinfo{year}{2017}).

\bibitem[{\citenamefont{Abdelsalam et~al.}(2016)\citenamefont{Abdelsalam,
  Talaat, Lukyanchuk, Portnoi, and Saroka}}]{abdelsalam2016electro}
\bibinfo{author}{\bibfnamefont{H.}~\bibnamefont{Abdelsalam}},
  \bibinfo{author}{\bibfnamefont{M.}~\bibnamefont{Talaat}},
  \bibinfo{author}{\bibfnamefont{I.}~\bibnamefont{Lukyanchuk}},
  \bibinfo{author}{\bibfnamefont{M.}~\bibnamefont{Portnoi}}, \bibnamefont{and}
  \bibinfo{author}{\bibfnamefont{V.}~\bibnamefont{Saroka}},
  \bibinfo{journal}{Journal of Applied Physics} \textbf{\bibinfo{volume}{120}},
  \bibinfo{pages}{014304} (\bibinfo{year}{2016}).

\bibitem[{\citenamefont{G{\"u}ttinger et~al.}(2012)\citenamefont{G{\"u}ttinger,
  Molitor, Stampfer, Schnez, Jacobsen, Dr{\"o}scher, Ihn, and
  Ensslin}}]{guttinger2012transport}
\bibinfo{author}{\bibfnamefont{J.}~\bibnamefont{G{\"u}ttinger}},
  \bibinfo{author}{\bibfnamefont{F.}~\bibnamefont{Molitor}},
  \bibinfo{author}{\bibfnamefont{C.}~\bibnamefont{Stampfer}},
  \bibinfo{author}{\bibfnamefont{S.}~\bibnamefont{Schnez}},
  \bibinfo{author}{\bibfnamefont{A.}~\bibnamefont{Jacobsen}},
  \bibinfo{author}{\bibfnamefont{S.}~\bibnamefont{Dr{\"o}scher}},
  \bibinfo{author}{\bibfnamefont{T.}~\bibnamefont{Ihn}}, \bibnamefont{and}
  \bibinfo{author}{\bibfnamefont{K.}~\bibnamefont{Ensslin}},
  \bibinfo{journal}{Reports on Progress in Physics}
  \textbf{\bibinfo{volume}{75}}, \bibinfo{pages}{126502}
  (\bibinfo{year}{2012}).

\bibitem[{\citenamefont{van~der Wiel et~al.}(2003)\citenamefont{van~der Wiel,
  De~Franceschi, Elzerman, Fujisawa, Tarucha, and
  Kouwenhoven}}]{Wiel2003RevModPhys}
\bibinfo{author}{\bibfnamefont{W.~G.} \bibnamefont{van~der Wiel}},
  \bibinfo{author}{\bibfnamefont{S.}~\bibnamefont{De~Franceschi}},
  \bibinfo{author}{\bibfnamefont{J.~M.} \bibnamefont{Elzerman}},
  \bibinfo{author}{\bibfnamefont{T.}~\bibnamefont{Fujisawa}},
  \bibinfo{author}{\bibfnamefont{S.}~\bibnamefont{Tarucha}}, \bibnamefont{and}
  \bibinfo{author}{\bibfnamefont{L.~P.} \bibnamefont{Kouwenhoven}},
  \bibinfo{journal}{Rev. Mod. Phys.} \textbf{\bibinfo{volume}{75}},
  \bibinfo{pages}{1} (\bibinfo{year}{2003}).

\bibitem[{\citenamefont{Abergel
  et~al.}(2010{\natexlab{b}})\citenamefont{Abergel, Apalkov, Berashevich,
  Ziegler, and Chakraborty}}]{Abergel10}
\bibinfo{author}{\bibfnamefont{D.~S.~L.} \bibnamefont{Abergel}},
  \bibinfo{author}{\bibfnamefont{V.}~\bibnamefont{Apalkov}},
  \bibinfo{author}{\bibfnamefont{J.}~\bibnamefont{Berashevich}},
  \bibinfo{author}{\bibfnamefont{K.}~\bibnamefont{Ziegler}}, \bibnamefont{and}
  \bibinfo{author}{\bibfnamefont{T.}~\bibnamefont{Chakraborty}},
  \bibinfo{journal}{Adv. Phys.} \textbf{\bibinfo{volume}{59}},
  \bibinfo{pages}{261} (\bibinfo{year}{2010}{\natexlab{b}}).

\bibitem[{\citenamefont{Ma and Li}(2012{\natexlab{b}})}]{Ma12}
\bibinfo{author}{\bibfnamefont{W.~L.} \bibnamefont{Ma}} \bibnamefont{and}
  \bibinfo{author}{\bibfnamefont{S.~S.} \bibnamefont{Li}},
  \bibinfo{journal}{Appl. Phys. Lett.} \textbf{\bibinfo{volume}{100}},
  \bibinfo{pages}{163109} (\bibinfo{year}{2012}{\natexlab{b}}).

\bibitem[{\citenamefont{Dong et~al.}(2007)\citenamefont{Dong, Liu, Teng, Zhang,
  and Cheng}}]{dong2007}
\bibinfo{author}{\bibfnamefont{Q.-R.} \bibnamefont{Dong}},
  \bibinfo{author}{\bibfnamefont{C.-X.} \bibnamefont{Liu}},
  \bibinfo{author}{\bibfnamefont{S.-Y.} \bibnamefont{Teng}},
  \bibinfo{author}{\bibfnamefont{N.-Y.} \bibnamefont{Zhang}}, \bibnamefont{and}
  \bibinfo{author}{\bibfnamefont{C.-F.} \bibnamefont{Cheng}},
  \bibinfo{journal}{Journal of Physics D: Applied Physics}
  \textbf{\bibinfo{volume}{40}}, \bibinfo{pages}{730} (\bibinfo{year}{2007}).

\bibitem[{\citenamefont{Elliott and Dawber}(1979)}]{elliott1979symmetry}
\bibinfo{author}{\bibfnamefont{J.}~\bibnamefont{Elliott}} \bibnamefont{and}
  \bibinfo{author}{\bibfnamefont{P.}~\bibnamefont{Dawber}},
  \emph{\bibinfo{title}{Symmetry in Physics, Volume 1 and 2}}
  (\bibinfo{publisher}{Macmillan}, \bibinfo{year}{1979}).

\bibitem[{\citenamefont{Dong}(2014)}]{dong2014electronic}
\bibinfo{author}{\bibfnamefont{Q.-R.} \bibnamefont{Dong}},
  \bibinfo{journal}{RSC Advances} \textbf{\bibinfo{volume}{4}},
  \bibinfo{pages}{12287} (\bibinfo{year}{2014}).

\end{thebibliography}
\end{document}